\documentclass[12pt]{iopart}
\usepackage{dcolumn}
\usepackage{bm}
\usepackage[dvips]{graphicx}

\begin{document} 

\title{Transport of Molecular Motor Dimers in Burnt-Bridge Models}
\author{Alexander Yu. Morozov and Anatoly B. Kolomeisky}
\address{Department of Chemistry, Rice University, Houston, TX 77005 USA}

\begin{abstract}
Dynamics of  molecular motor dimers, consisting of rigidly bound particles that move along two parallel lattices and interact with underlying molecular tracks, is investigated theoretically by analyzing discrete-state stochastic continuous-time burnt-bridge models. In these models the motion of molecular motors is viewed as a random walk along the lattices with periodically distributed  weak links (bridges). When the particle crosses the weak link it can be destroyed with a probability $p$, driving the molecular motor motion in one direction.  Dynamic properties and effective generated forces of dimer molecular motors are calculated exactly as a function of a concentration of bridges $c$ and burning probability $p$ and compared with properties of the monomer motors. It is found that the ratio of the velocities of the dimer and the monomer can never exceed 2, while the dispersions of the dimer and the monomer are not very different. The relative effective generated force of the dimer (as compared to the monomer)  also  cannot be larger than 2  for most sets of parameters. However,  a very large force can be produced by the dimer in the special case of $c=1/2$ for non-zero shift between the lattices. Our calculations do not show the significant increase in the force generated by collagenase motor proteins in real biological systems as predicted by previous computational studies. The observed behavior of dimer molecular motors is discussed by considering in detail the particle dynamics near burnt bridges.
\end{abstract}

\ead{tolya@rice.edu}

\maketitle

\section{Introduction}

Theoretical and experimental investigations of motor proteins, or molecular motors, have recently become a subject of considerable attention because of  importance of these  molecules  for fundamental understanding of non-equilibrium processes in chemistry and biology \cite{lodish_book,howard_book,bray_book,AR}. Motor proteins are active enzymatic molecules that consume a chemical energy and transform it into mechanical work, while moving at the same time along linear molecular tracks. They play a critical role in many biological processes \cite{lodish_book,howard_book,bray_book}. In contrast to conventional molecular motors powered by the hydrolysis of ATP or related compounds, a recently discovered molecular motor called collagenase fuels its motion along collagen fibrils by using a significantly different mechanism that utilizes asymmetric collagen proteolysis \cite{saffarian04,saffarian06}. The biased diffusion of this motor protein along the molecular track results from the fact that the motor protein  molecule after cutting the bond at the specific  site of the collagen fiber  is always found on one side of the reaction site, and it is unable to cross the broken link after the cleavage.

Analysis of collagenase transport  suggests that a successful description of this motor protein's dynamics  can be obtained by utilizing  so-called ``burnt-bridge models''  (BBM) \cite{saffarian04,saffarian06,mai01,antal05, morozov07,artyomov07}. According to  BBM, the molecular motor is described as a random walker hopping along the one-dimensional lattice composed of two kinds of links: strong and weak. The walker does not affect the strong links when crossing them, while the weak links (``bridges'') might be destroyed (``burnt'') with a probability $0<p\le 1$ when crossed by the walker; the walker cannot pass over the destroyed links again. In Refs. \cite{morozov07,artyomov07} we developed a new analytical approach that allowed us to obtain  exact expressions for motor protein velocity $V(c,p)$ and dispersion $D(c,p)$ for different sets of parameters and for different burnt-bridge models, and it was shown that the analytical predictions are in full agreement with Monte-Carlo  computer simulations.

Theoretical and experimental investigations of collagenase dynamics \cite{saffarian04,morozov07} conclude that this motor protein is rather weak: at standard conditions it exerts a stall force of  $F_{S} \simeq$ 0.02 pN, which is 100 times smaller than other conventional motor proteins \cite{AR}, and thermodynamic efficiency of this motor protein is between 6 and 15$\%$, although it has a significantly larger processivity  than conventional motor proteins \cite{saffarian04,morozov07}. Meantime, experimental studies of several collagenases capable of breaking collagen fibers \cite{saffarian06,itoh01,rozanov01} indicate that in cells these enzymes are probably clustered together. Computer simulations of the motion of rigidly bound dimers \cite{saffarian06} of collagenase molecules produced stall forces of the order of 5 pN, 10-100 times larger than the single motor protein, and it was suggested that cell surface-anchored clusters of collagenases might play an important role in cell motility. Coupling of motor proteins can produce very efficient molecular motors and it can lead to interesting dynamic phenomena \cite{stukalin05,stukalin06}. However, the motion of dimer particles in the context of BBM has not been studied yet. The goal of the present  work is to investigate theoretically the dynamics of dimer molecular motors in BBM using previously developed exact analytical approaches \cite{morozov07,artyomov07}.

\section{Model}

We  consider a molecular motor dimer as two particles connected by a rigid bond  that move along two parallel infinite one-dimensional lattices (a bond between the particles is perpendicular to the linear molecular tracks), with transition rates in both directions and the lattice spacing assumed to be equal to one, as shown in Fig. 1. The links connecting consecutive binding sites on the lattices fall in two categories. While the random walkers have no effect on the strong links, each of them destroys the weak ones with the probability $p$ when passing over them. After the weak link is broken, the particle is always assumed to be to the right of it, and all weak links  are intact at $t=0$. Therefore, the dimer cannot be trapped between two burnt bridges, and it continuously moves to the right.

\begin{figure}[ht]
\centering
\includegraphics[scale=0.6,clip=true]{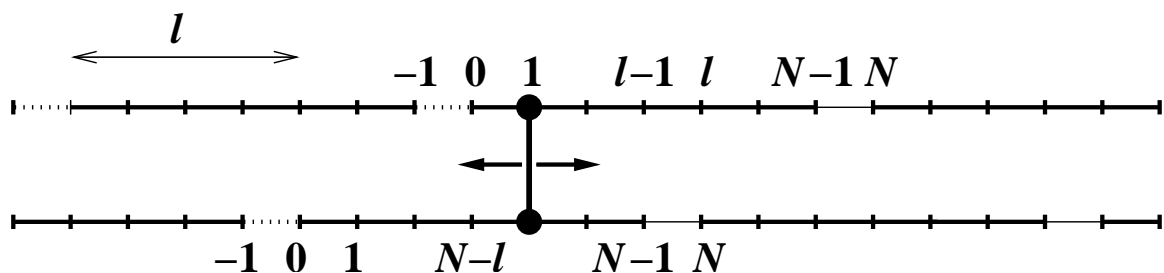}
\caption{Schematic view of the motion of molecular motor dimer in the burnt-bridge model. Strong links are represented by thick solid lines, while thin solid lines depict the weak links which can be destroyed. Dotted lines correspond to already burnt bridges. The dimer can jump with equal transition rates one step to the right or to the left provided that the link is not already destroyed. The bridges on the upper and lower tracks are shifted by $l$ lattice spacings.}
\end{figure}

The bridge can be burnt according to one of the two scenarios. In the first mechanism, the bridge burns only when the random walker crosses it from left to right, however it always remains intact if crossed from right to left. We term this model a forward BBM. According to the second model, the weak link gets broken if crossed in either direction, although the particle is always found to the right of the broken bond. This is a forward-backward BBM. The physics of the driven motion is similar for both mechanisms, and both models are even  identical for $p=1$, however for $p<1$ the walker's dynamics in two cases are  different \cite{morozov07}. In this paper we will  consider only  forward BBM because theoretical analysis is simpler in this case. However, our approach can be easily extended to more complex forward-backward BBM.

The dynamics of the dimer in BBM depends on two parameters: the burning probability $p$ and the concentration of bridges $c$. The properties of the system are also strongly influenced by the distribution of bridges, periodic or random  \cite{antal05}. In the case of the periodic distribution, which will be analyzed in detail in the present paper, the bridges on each of two parallel one-dimensional lattices are positioned equidistantly, with  a constant distance $N=1/c$ lattice spacings separating them. This distribution is more suited for the analysis of the transport of collagenases since the sites that can be destroyed by enzymes on collagen fibers are found to be at equal distances from each other \cite{saffarian04}.    All the models presented  below will be studied in the continuous-time description that  better corresponds to the real biological systems with motor proteins \cite{morozov07}. We also assume that the bridges on two parallel tracks are shifted by $l$ lattice spacings ($0 \le l \le N-1$): see Fig. 1.

\section{Dynamics of Molecular Motor Dimers}

\subsection{BBM with $p=1$}

\begin{figure}[ht]
\centering
\includegraphics[scale=0.6,clip=true]{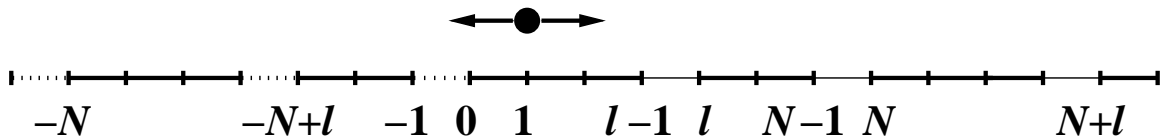}
\caption{Schematic picture for the transport of the single particle (equivalent to the dimer in BBM with $p=1$) along the effective single  linear track.}
\end{figure}

Let us consider first a special case of BBM with $p=1$, i.e., when a bridge burns every time when crossed by the walker. It can be  easily seen  that in this case  the motion of a molecular motor dimer along two parallel tracks  is equivalent to the motion of a single random walker on a single track but with increased concentration of bridges \cite{saffarian06}, as shown in Fig. 2. The distribution of bridges on this effective track is again periodic but generally non-uniform, with two weak links in each period: between the sites $l-1$ and $l$, and between the sites $N-1$ and $N$: see Fig.2. The dynamics of this system can be mapped into a random single-particle hopping model on periodic lattices  that was solved by Derrida \cite{AR,derrida83}, in which the explicit expressions for the dynamic properties are obtained. We define the forward and backward transition rates for the random walker at site $j$ ($j=0,1,\cdots,N-1$) as  $u_j$ and $w_j$, correspondingly. The velocity and the diffusion constant are then given in terms of these transition rates \cite{AR,derrida83},
\begin{equation}\label{vel}
V =\frac{N}{\left(\sum\limits_{j=0}^{N-1}r_j\right)} \left[1-\prod\limits_{i=1}^{N}\left( \frac{w_i}{u_i}\right)\right],
\end{equation}
\begin{equation}\label{D}
D = \frac{1}{\left(\sum\limits_{j=0}^{N-1}r_j\right)^2} \left\{ V \sum\limits_{j=0}^{N-1} s_j\sum\limits_{i=0}^{N-1} (i+1)r_{j+i+1} +N \sum\limits_{j=0}^{N-1} u_j s_j r_j \right\} -\frac{V (N+2)}{2},
\end{equation}
with auxiliary functions $r_j$ and $s_j$ defined as
\begin{equation}\label{r_j}
r_j = \frac{1}{u_j}\left[ 1+\sum\limits_{k=1}^{N-1} \prod\limits_{i=j+1}^{j+k}\left(\frac{w_i}{u_i}\right)\right],
\end{equation}
and
\begin{equation}\label{s_j}
s_j = \frac{1}{u_j}\left[ 1+\sum\limits_{k=1}^{N-1} \prod\limits_{i=j-1}^{j-k}\left(\frac{w_{i+1}}{u_i}\right)\right].
\end{equation}

The transition rates for the particle shown in Fig. 2 are $u_{j}=w_{j}=1$  for all $j$ except $w_0=w_l=0$ (the burnt bridges cannot be crossed again). Then the explicit expressions for (\ref{r_j}) and (\ref{s_j}) are the following,
\begin{equation}
r_j =\left\{ \begin{array}
{r@{\quad \quad}l} l-j, & \mbox{ for } j=0,1,\cdots,l-1; \\ 
                   N-j, & \mbox{ for } j=l,l+1,\cdots,N-1;  
\end{array} \right. 
\end{equation}
and
\begin{equation}
s_j =\left\{ \begin{array}
{r@{\quad \quad}l}  j+1, & \mbox{ for } j=0,1,\cdots,l-1; \\
                    j-l+1, & \mbox{ for } j=l,l+1,\cdots,N-1. 
\end{array} \right. 
\end{equation}
Substituting these results into Eq.(\ref{vel}) leads to the expression for the dimer's velocity,
\begin{equation}\label{vel1}
V(l,N)=\frac{2N}{l(l+1)+(N-l)(N-l+1)}.
\end{equation}

Calculation of the dispersion $D$ according to (\ref{D}) is more complex, but it can be performed with the help of the following  useful identities:
\begin{equation}
r_{j+i+1} =\left\{ \begin{array}
                   {r @{\quad \quad}l} l-(j+i+1), & \mbox{ for } 0\le i\le l-j-2; \\ 
                   N-(j+i+1), & \mbox{ for } l-j-1\le i\le N-j-2; \\ 
                   l+N-(j+i+1), & \mbox{ for } N-j-1\le i\le N-1, 
\end{array} \right. 
\end{equation}
for $j\le l-1$, while for $j\ge l$ we have
\begin{equation}
r_{j+i+1} =\left\{ \begin{array}
                 {r@{\quad \quad}l} N-(j+i+1), & \mbox{ for } 0\le i\le N-j-2; \\ 
                 l+N-(j+i+1), & \mbox{ for } N-j-1\le i\le N+l-j-2; \\ 
                 2N-(j+i+1), & \mbox{ for } N+l-j-1\le i\le N-1. 
\end{array} \right. 
\end{equation}
As a result, the dispersion for the dimer is given by the following formula,
\begin{eqnarray}\label{D1}
D(l,N) & = & \frac{2N^2}{3(2l^2+N-2lN+N^2)^3} \left[ 2l^4-4l^3N-2lN(2+3N+2N^2) \right. \nonumber \\
  & & \left. +2 l^2(2+3N+3N^2)+N(1+N)(1+N+N^2) \right].
\end{eqnarray}

The validity of Eqs. (\ref{vel1}) and (\ref{D1}) is confirmed by the fact that for $l=0$ and $l=N$ they reproduce the BBM results obtained for the transport of monomers on the lattice with the  period $N$ \cite{morozov07},
\begin{equation}\label{vd1}
 V=\frac{2}{N+1}, \ \ \  D=\frac{2(N^2+N+1)}{3(N+1)^2}; 
\end{equation}
whereas for $N=2l$ we recover the results for the monomer's dynamics on the lattice with the period $l$, as expected. Also, Eqs. (\ref{vel1}) and (\ref{D1}) are invariant under the change $l \rightarrow N-l$, reflecting the symmetry of the system.

For the purposes of comparison  of the dynamic properties of the  dimer with those of the monomer in BBM with $p=1$, we plot their ratios in Figs. 3 and 4 as functions of the concentration of bridges $c=1/N$ and the shift $l$ between the parallel tracks.  The range of $l$ values (for constant $c$) is $0\le l\le N=1/c$, whereas that of $c$ values (for constant $l$) is $0\le c\le 1/l$. Hence it is convenient to express the dynamic properties of the dimer as functions of dimensionless parameter $lc$ (with either fixed $l$ or $c$), so that $0\le lc\le1$ and the curves with different $l$ (or $c$) values could be plotted on the same interval.  The physical meaning of the parameter $lc$ is a dimensionless concentration of bridges (for fixed $l$), or a dimensionless shift distance (for fixed $c$). The invariance of dynamic properties (\ref{vel1}) and (\ref{D1}) under replacement $l \rightarrow N-l$ implies that when plotted as functions of $l$ (for fixed $c$), their graphs are symmetric with respect to the middle of the $lc$ interval, i.e. $lc=\frac{1}{2}$: see Figs 3b and 4b.

\begin{figure}[tbp]
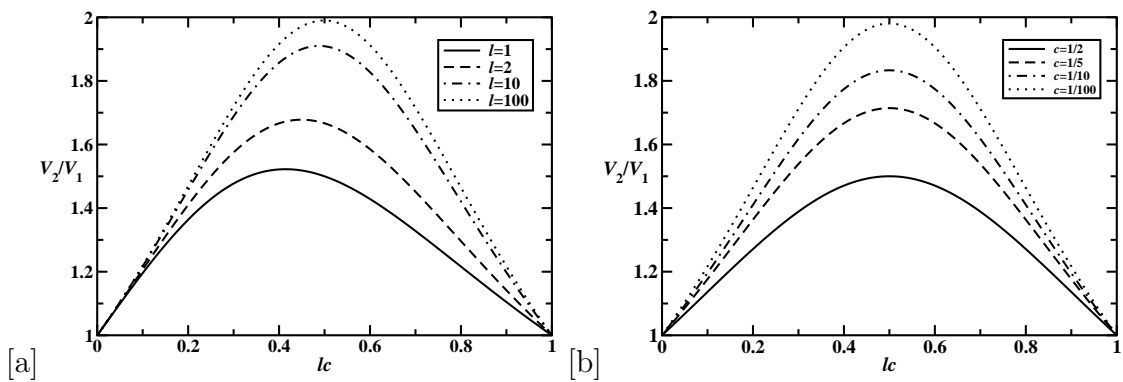

\centering
[a]\includegraphics[scale=0.28,clip=true]{Fig3a.eps}
[b]\includegraphics[scale=0.28,clip=true]{Fig3b.eps}
\caption{(a) Ratio of dimer's and monomer's velocities as a function of the concentration of bridges $c$ for the fixed value of the shift $l$.  (b) Ratio of dimer's and monomer's velocities as a function of the shift $l$ for the fixed concentration of bridges $c$.}
\end{figure}

\begin{figure}[tbp]
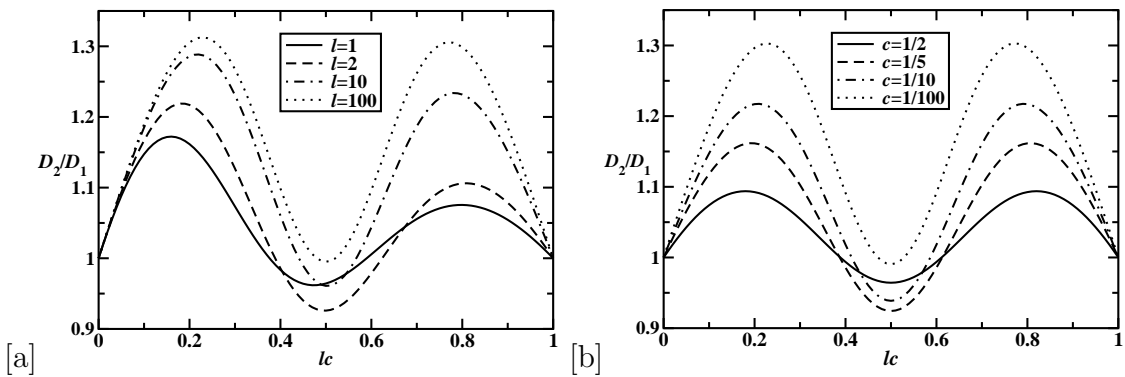

\centering
[a]\includegraphics[scale=0.28,clip=true]{Fig4a.eps}
[b]\includegraphics[scale=0.28,clip=true]{Fig4b.eps}
\caption{ (a) Ratio of dimer's and monomer's dispersions as a function of the concentration of bridges for the fixed value of the shift $l$.  (b) Ratio of dimer's and monomer's dispersions as a function of the shift $l$ for the fixed concentration of bridges $c$.}
\end{figure}

As shown in Fig. 3a, the relative velocity of the dimer has a non-monotonic behavior as a function of the concentration of weak links, and $V_{2}/V_{1}$ reaches a maximum at $c^{\ast}=\frac{1}{l+\sqrt{l^2+l}}$, at which $V_2(l,c^{\ast})/V_1(c^{\ast})=1+\frac{2l}{2\sqrt{l^2+l}+1}<2$. For very large shifts, $l \gg 1$, one can easily see that  $c^{\ast}\rightarrow\frac{1}{2l}$ and $V_2(l,c^{\ast})/V_1(c^{\ast})\simeq 2-\frac{1}{l}$. Similar behavior is observed for the relative velocity of the dimer as a function of the shift distance $l$, as illustrated in Fig. 3b. As discussed above, because of the symmetry the maximum of $V_2/V_1$ is observed at $l^{\ast}=\frac{N}{2}=\frac{1}{2c}$, at which $V_2(l^{\ast},c)/V_1(c)=\frac{2(N+1)}{N+2}$, and for very low concentration of bridges ($N \rightarrow \infty$) the dimer molecule moves twice as fast as the monomer molecular motor at these conditions. Thus, we conclude that at all conditions  $V_2(l,c)/V_1(c) \le 2$. It is also worth mentioning that the dimer's velocity itself $V_2(l,c)$ reaches the maximum at $\tilde{c}=\frac{1}{\sqrt{2}l}$ (for fixed $l$) and $\tilde{l}=l^{\ast}=\frac{N}{2}=\frac{1}{2c}$ (for fixed $c$), as can be deduced from Eq. (\ref{vel1}). The  corresponding values of $V_2$  are $V_2(l,\tilde{c})=\frac{2}{2(\sqrt{2}-1)l+1}$ and $V_2(\tilde{l},c)=\frac{4}{N+2}$.

The  dispersion of the dimer shows a more complex behavior with two maxima and one minimum, as presented in Fig. 4. Our calculations show that generally $D_{2}/D_{1}$ lies between 0.92 and 1.31. It is interesting to note that for some range of parameters (not very large shifts and not very small concentration of bridges) the dimer fluctuates less than the monomer, leading to $D_{2}/D_{1}<1$.

Simultaneous knowledge of velocities and dispersions of the dimer allows us to evaluate the effective force exerted by the molecular motor \cite{morozov07}. The motion of the dimer in the system with burnt bridges can be viewed as an effective biased random walk in the system without bridges. Then its dynamic properties can also be  written as
\begin{equation}\label{}
\tilde{V}=a(u_{eff}-w_{eff}), \quad  \tilde{D}=\frac{1}{2}a^2(u_{eff}+w_{eff}),
\end{equation}
with $u_{eff}$ and $w_{eff}$ being the effective transition rates of the biased random walk \cite{morozov07}. The effective force generated by the dimer is given by \cite{AR,fisher99}
\begin{equation}\label{Fs}
F=\frac{k_B T}{a}\ln \frac{u_{eff}}{w_{eff}}=\frac{k_B T}{a}\ln \frac{2\tilde{D}+\tilde{V}a}{2\tilde{D}-\tilde{V}a},
\end{equation}
where $a$ is the size of the discrete steps made by the walker, and it corresponds to the distance between the consecutive lattice sites. Parameters $\tilde{V}$ and $\tilde{D}$ are expressed in units of $m/s$ and $m^2/s$ correspondingly, and they are related to the dimensionless dynamic properties $V$ and $D$ [Eqs. (\ref{vel1}) and (\ref{D1})] in the following way: 
\begin{equation}\label{}
\tilde{V}=a v_0 V, \quad \tilde{D}=a^2 v_0 D,
\end{equation}
where $v_0$ is the intrinsic transition rate of the molecular motor. Then
\begin{equation}\label{f2}
F(l,N)=\frac{k_B T}{a}\ln \frac{2D+V}{2D-V},
\end{equation}
with functions  $V(l,N)$ and $D(l,N)$  given by Eqs. (\ref{vel1}) and (\ref{D1}) respectively.

To investigate the efficiency of the dimer molecular motor we compare the force generated by the dimer particle  in BBM with the force exerted by the monomer at the same conditions, and results are presented in Fig. 5. Note that $F_2(l,c)$ is given by Eq. (\ref{f2}), while the corresponding force for the monomer is equal to $F_1(c)=F_2(l=0,c)$. As shown in Fig. 5a, for $l \ge 2$ the ratio of effective forces is less than 2 for all concentrations of bridges, although the maximum of the ratio might slightly exceed 2 for small $l$ ($\sim 2.04$ for $l=2$). The situation is drastically different for $l=1$ where $F_2/F_1\rightarrow\infty$ as $c\rightarrow 1/2$ (since $\lim\limits_{c\rightarrow 1/2}^{}F_2=\infty$ for $l=1$). For all curves in Fig. 5a, the maximum  is reached very close to $lc=0.5$, and it approaches   $lc=0.5$ as $l\rightarrow\infty$. For fixed concentration of bridges (see Fig. 5b), symmetry requires that $F_2(N-l,c)=F_2(l,c)$, and $F_2/F_1$ reaches its maximum at $l^{\star}=\frac{N}{2}=\frac{1}{2c}$. It is clear from Fig. 5b that except for the case of $c=1/2$, where $F_2/F_1\rightarrow\infty$ as $l\rightarrow 1$, the maximum of the ratio $F_2/F_1$  is never larger than 2. It approaches 2 (from below) as $N\rightarrow\infty$ ($c\rightarrow 0$).

\begin{figure}[tbp]
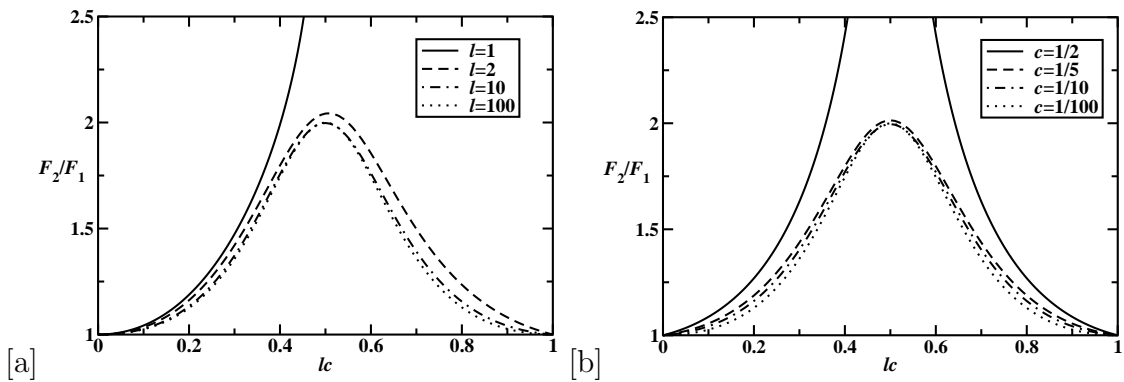

\centering
[a]\includegraphics[scale=0.28,clip=true]{Fig5a.eps}
[b]\includegraphics[scale=0.28,clip=true]{Fig5b.eps}
\caption{(a) Ratio of dimer's and monomer's effective forces as a function of the concentration of bridges $c$ for the fixed value of the shift $l$. Note that for $l=1$ there is no branch for $lc>1/2$ because there are no possible values of $c$ between 1 and 1/2 ($N$ is integer). (b) Ratio of dimer's and monomer's effective forces as a function of the shift $l$ for the fixed concentration of bridges $c$.}
\end{figure}

\subsection{BBM with $p\le 1$: a special case}

Although the general situation when the burning probability is $p<1$ requires a separate consideration that leads to  extensive calculations (see next section), there is a special case of BBM for which the dynamics of the dimer molecular motor can be  expressed in terms of the properties of the effective monomer in BBM, for which analytical expressions are already known \cite{morozov07,artyomov07}. When bridges on the parallel tracks are shifted by the distance $l=N/2$, i.e., half of the period, the motion of the dimer is identical to hopping of the effective particle on the single lattice with the distance of $N/2$ between neighboring weak links. Then, using results from Ref. \cite{morozov07,artyomov07}, the relations between the dynamic properties of the dimer and the corresponding monomer are the following for $l=N/2$,
\begin{equation}\label{vd}
V_2(c,p)=V_1(2c,p), \quad  D_2(c,p)=D_1(2c,p).
\end{equation}

To simplify calculations we only consider  forward BBM, in which bridges burn only when crossed from left to right, but remain unaffected if crossed in the opposite direction. For this model, analytical expressions for $V_1(c,p)$ and $D_1(c,p)$ are already obtained  \cite{morozov07,artyomov07}. The velocity of the monomer is 
\begin{equation}\label{v1}
V_1(c,p)=\frac{2pc}{\sqrt{p^2(1-c)^2 + 4pc}}, 
\end{equation}
while the dispersion has more complex form \cite{artyomov07},
\begin{eqnarray} \label{finalD}
D_1=\frac{1}{2}\left[-2\Delta+2-R_0-2R_0 \left \{\frac{Ax}{2(1-x)^2}[N+1+(N-1)x]-\frac{N(N^2-1)}{6(1-x)}\Gamma \right. \right. \nonumber \\
\left. \left. -\frac{(N-2)(N^2-1)}{24}R_0^2-\frac{N}{1-x}\Delta \right \} \right]
\end{eqnarray}
where parameters $A$ and  $\Delta$ are given by 
\begin{eqnarray}\label{A}
A &=&\frac{N(1-x)[\alpha px+R_0(1-x)(-1+p+x)]}{p[1+(N-1)x](-1+p+x^2)}, \\
\label{delta}
\Delta &=&\frac{(1-p)[\alpha p+R_0(1-x)^2]}{p(-1+p+x^2)}
\end{eqnarray} 
with 
\begin{equation}\label{alpha}
\alpha =\frac{N(N-1)}{2}\Gamma +\frac{(N-1)(N-2)}{6}(1-x)R_0^2,
\end{equation} 
and
\begin{equation}\label{gamma}
\Gamma =\frac{2R_0}{N}(1-x)-R_0^2.
\end{equation}
Thus Eq. (\ref{finalD}) is expressed in terms of the parameters $R_0$ and $x$ which are also functions of $N=1/c$ and $p$,
\begin{equation}\label{rr0}
R_0 = \frac{2p}{\sqrt{p^2(N-1)^2 + 4pN}} = \frac{2pc}{\sqrt{p^2(1-c)^2 + 4pc}},
\end{equation}
\begin{equation}\label{x}
x = 1+\frac{1}{2}p(N-1)-\frac{1}{2}\sqrt{p^2(N-1)^2 + 4pN}.
\end{equation}
Therefore, Eq. (\ref{finalD}) combined with Eqs. (\ref{A}) - (\ref{x}) provides the explicit expression for $D_1(c,p)$ as a function of the concentration of bridges and the bridge burning probability.

\begin{figure}[tbp]
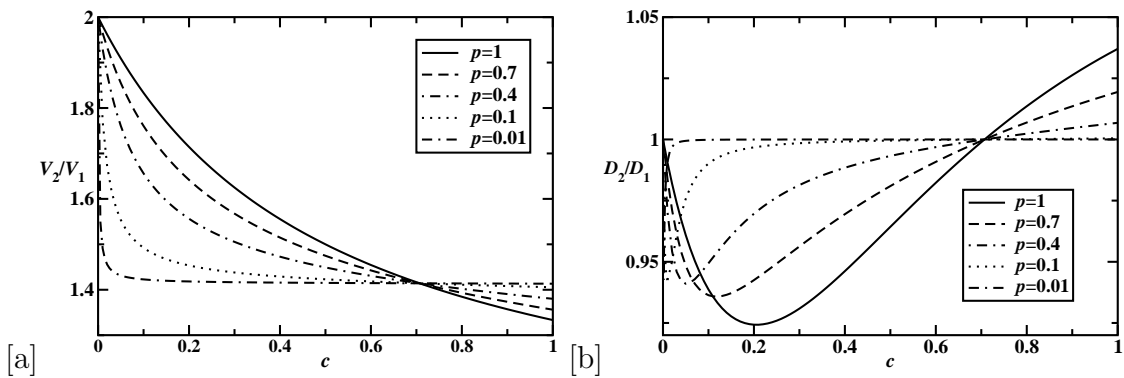
\label{fig6}
\centering
[a]\includegraphics[scale=0.28,clip=true]{Fig6a.eps}
[b]\includegraphics[scale=0.28,clip=true]{Fig6b.eps}
\caption{(a). Ratio of dimer's and monomer's velocities as a function of concentration of bridges for $l=1/(2c)$.  (b) Ratio of dimer's and monomer's dispersions as a function of concentration of bridges for $l=1/(2c)$.}
\end{figure}

In this case the dynamic properties of dimer and monomer molecular motors can be compared  for the entire range of burning probabilities, $0 < p \le 1$, as presented in Fig. 6. Since  the shift $l$ assumes only integer values ($l=1,2,\cdots$), the relation $lc=1/2$ implies that we only need to consider the set of discrete $c$ values, $c=1/2,1/4,1/6,\cdots$, i.e., only the interval $0 < c \le 1/2$ is relevant. As seen from Fig. 6a, decreasing the burning probability for any fixed concentration of bridges lowers the relative velocity of the dimer. Increasing the number of weak links (for fixed $p$) also lowers the relative velocity of the dimer, and the ratio of velocities lies within the interval, $\sqrt{2}<V_2/V_1\le 2$. This result can be explained using the asymptotic expressions for the velocity  obtained earlier \cite{artyomov07},
\begin{equation}\label{vlim}
V_1(c,p) \simeq \left\{ \begin{array}
                        {r@{\quad \quad}l}  2c, & \mbox{ for } c\ll p \\ 
                                    \sqrt{pc}, & \mbox{ for } p\ll c.  
\end{array} \right. 
\end{equation}
Then for $c\rightarrow 0$ the ratio $V_{2}/V_{1}$ approaches 2 for any finite value of the burning probability, while for $p \ll 1$ (for finite $c$) we have $V_{2}/V_{1} \simeq \sqrt{2}$. As shown in Fig. 6b, the dispersion of the dimer is similar to the dispersion of the monomer, $0.92 \le D_2/D_1\leq 1$ for $0<c\le 1/2$. Lowering the burning probability increases the relative diffusion constant of the dimer particle, while the effect of the bridge concentration is non-monotonic.  Because in the limiting cases the dispersion is independent of the concentration of bridges \cite{morozov07} [$D_{1}(c,p) \simeq 2/3$ for $c \ll p$ and $D_{1}(c,p) \simeq 1/2$ for $p \ll c$], the ratio of dispersion is  $D_2/D_1\rightarrow 1$ in these cases.

The effective forces exerted by dimer and monomer molecular motors for $p \le 1$ and $lc=1/2$ can be obtained using the approach described in the previous section for the case of $p=1$, and the results are plotted in Fig. 7.  The force exerted by a single particle $F_1(c,p)$ is found from Eq. (\ref{f2}) with $D=D_1(c,p)$ and $V=V_1(c,p)$ given by Eqs. (\ref{finalD}) and (\ref{v1}). Similarly, the force generated by a dimer $F_2(c,p)$ is found from Eq. (\ref{f2}) with $D=D_2(c,p)$ and $V=V_2(c,p)$, and thus $F_2(c,p)=F_1(2c,p)$.  Making use of asymptotic relations for the velocity and dispersion \cite{artyomov07}  yields the following expressions for $F_1$ in the limiting cases,
\begin{equation}\label{Flim}
F_1(c,p) \simeq \left\{ \begin{array}
{r@{\quad \quad}l} 3c, & \mbox{ for } c\ll p; \\ 
           2\sqrt{pc}, & \mbox{ for } p\ll c.  
\end{array} \right. 
\end{equation}
It follows that $F_2/F_1 \rightarrow 2$ if $c\rightarrow 0$ for any finite value of $p$, and  $F_2/F_1 \rightarrow\sqrt{2}$ for small burning probabilities (with fixed value of $c$). For $p=1$, the ratio $F_{2}/F_{1}$ strongly increases as $c \rightarrow 1/2$. For $p<1$, each  curve shows a non-monotonic behavior with a minimum. The position of the minimum moves to the right with decreasing $p$, and it moves beyond the interval $0<c\le 1/2$ for $p\le 0.2$. The minimal value of $F_2/F_1$ decreases with decreasing burning probability. The ratio $F_2/F_1$ exceeds 2 within the interval $0<c\le 1/2$ if $p \ge  0.74$, this occurs in the vicinity of $c=1/2$. In particular, for $p$ close to 1 the ratio $F_2/F_1$ at $c=1/2$ is substantially larger than 2 [e.g. at $c=1/2$, $F_2/F_1\approx 2.52$ for $p=0.9$, $F_2/F_1\approx 3.94$ for $p=0.99$, $F_2/F_1\approx 5.44$ for $p=0.999$].  Our analysis indicates that the relative force exerted by the dimer could substantially exceed 2 only at $c=1/2$ if the burning probability $p$ is close to 1.

\begin{figure}[tbp]
\centering
\includegraphics[scale=0.28,clip=true]{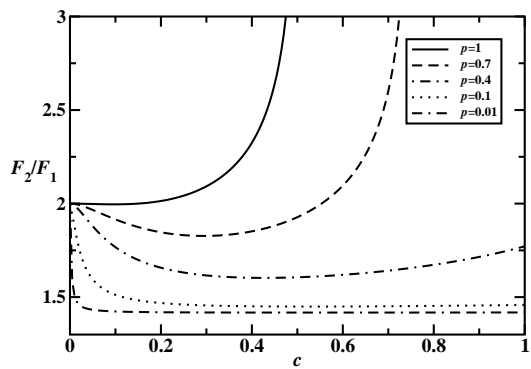}
\caption{Ratio of the force exerted by dimer and that exerted by monomer as a function of concentration of bridges $c$ with shift $l=1/(2c)$.}
\end{figure}

Given that in the $p=1$ case with fixed $c$ the ratios $V_2/V_1$ and $F_2/F_1$ reach their maximal values at $l=N/2$ (as can be seen from Figs. 3b and 5b), it is reasonable to suggest that the same holds for $p<1$. Thus the special case of $lc=\frac{1}{2}$ considered in the present section provides the maximal values for ratios $V_2/V_1$ and $F_2/F_1$ for any given $c$, i.e. for $l$ values other than $l=\frac{1}{2c}$ these ratios would be smaller.

\subsection{BBM with $p\le 1$: general case}

Now let us  consider the general case of the transport of the dimer molecular motors along two parallel tracks shifted by $l$ lattice spacings with $l=0,1,2,\cdots,N-1$ and with the burning probability $p\le 1$ (see Fig. 1). Again, we  analyze the continuous-time forward BBM when the bridges are destroyed only when crossed from left to right. For  burning probability $p<1$ the Derrida's method is not applicable since we do not know how the transition rates $u_j$ and $w_j$ are expressed in terms of $p$ \cite{morozov07}. Instead, we will apply  a method  developed  in Ref. \cite{morozov07} to investigate  BBM with $p<1$ for the monomer random walkers.  It should be pointed out, however, that this approach allows us to calculate only velocities. To determine the diffusion constants one should utilize a more complex approach outlined in Ref. \cite{artyomov07}, but in this case the calculations become very involved and we will not discuss them here.

\begin{figure}[ht]
\centering
\includegraphics[scale=0.5,clip=true]{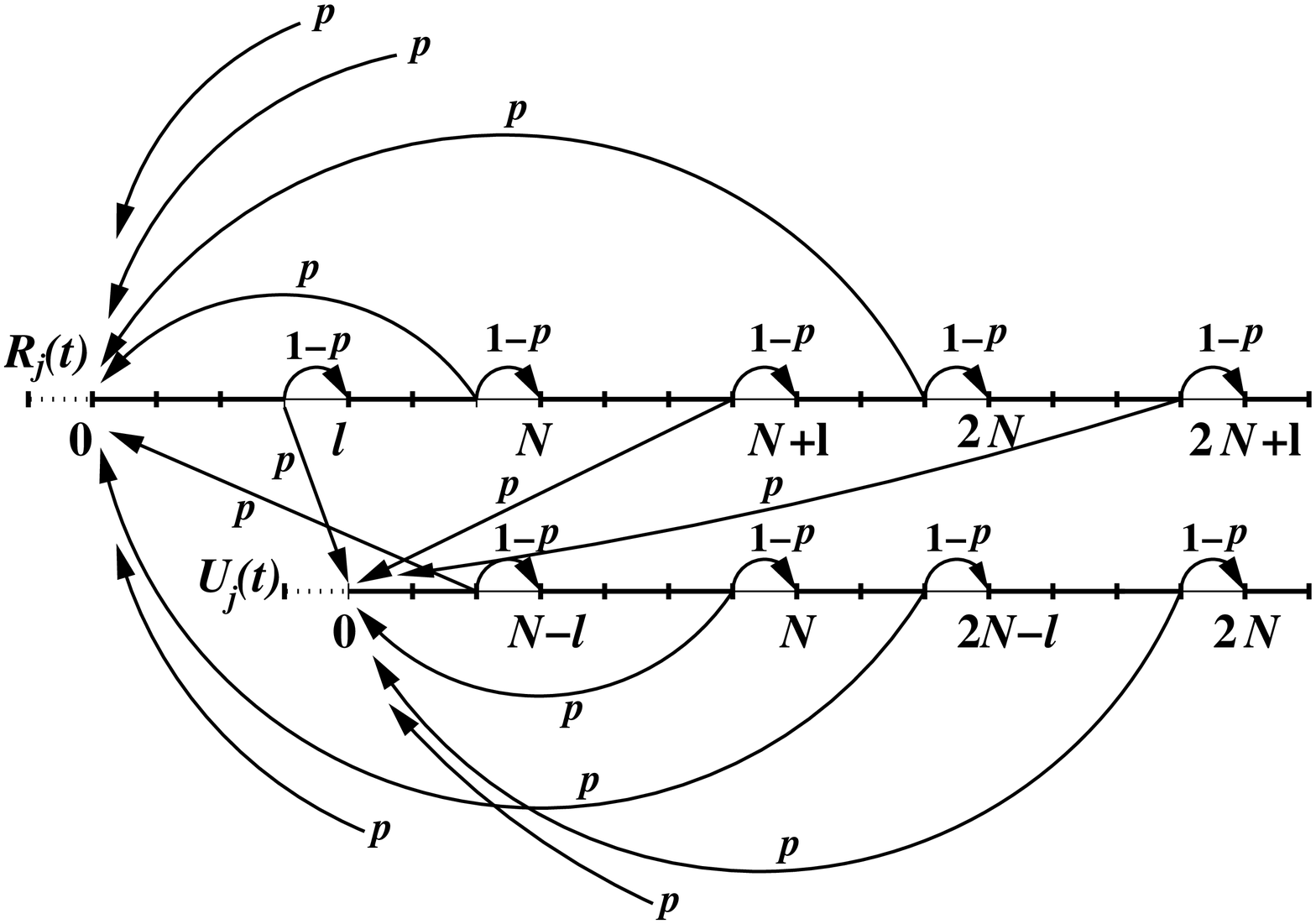}
\caption{Reduced kinetic scheme for the motion of the dimer in continuous-time forward BBM with bridges on the parallel tracks shifted by $l$ lattice spacings. Only transition rates not equal to one are shown. Two origins correspond to the right end of the last burnt bridge at each track.}
\end{figure}

Let us define $R_j(t)$ [$U_j(t)$] as the probability that a random walker is located $j$ sites apart from the last burnt bridge at time $t$ given that the last burnt bridge lies on the upper (lower) of the two tracks (see Fig. 1). The probabilities $R_j(t)$ and $U_j(t)$ are conveniently described  in a reduced kinetic scheme presented in Fig. 8. The description of the dimer's dynamics in terms of the functions $R_j(t)$ and $U_j(t)$ corresponds to considering the system in the moving coordinate frame with the last burnt bridge always at the origin of the upper or lower linear track. The kinetic scheme in Fig. 8 describes correctly  the dynamics for $l=1,2,\cdots,N-1$. However, the case of zero shift, $l=0$,  requires a slightly different analysis, which will be given below.

The time evolution of the system is determined by a set of Master equations. For the sites on the upper track the equations are
\begin{equation}\label{me1}
\frac{d R_{kN+i}(t)}{dt} = R_{kN+i-1}(t)+R_{kN+i+1}(t)-2 R_{kN+i}(t),
\end{equation}
for $k=0,1,2,\cdots$ and $1 \le i \le N-1$ (but with $i \ne l$). For the sites with $i=l$ we have 
\begin{equation}\label{me2}
\frac{d R_{kN+l}(t)}{dt}=(1-p)R_{kN+l-1}(t)+R_{kN+l+1}(t)-2R_{kN+l}(t);
\end{equation}
while for $i=0$ and $k \ge 1$ 
\begin{equation}\label{me3}
\frac{d R_{kN}(t)}{dt}=(1-p)R_{kN-1}(t)+R_{kN+1}(t)-2R_{kN}(t).
\end{equation}
For the origin site the dynamics is more complex,
\begin{equation}\label{me4}
\frac{d R_{0}(t)}{d t}=p\left[\sum_{k=1}^{\infty} R_{kN-1}(t)+\sum_{k=1}^{\infty} U_{kN-l-1}(t)\right] +R_{1}(t)-R_{0}(t).
\end{equation}
The Master equations for $U_j(t)$ are obtained from (\ref{me1}) - (\ref{me4}) by changing $R(t)\rightarrow U(t)$ and  $l\rightarrow N-l$. Specifically, at the origin site of the lower track we obtain  
\begin{equation}\label{me5}
\frac{d U_{0}(t)}{d t}=p\left[\sum_{k=1}^{\infty} U_{kN-1}(t)+\sum_{k=0}^{\infty} R_{kN+l-1}(t)\right] +U_{1}(t)-U_{0}(t).
\end{equation}
The probabilities $R_j(t)$ and $U_j(t)$ should also satisfy a normalization condition,
\begin{equation}\label{norm}
\sum_{k=0}^{\infty} \sum_{i=0}^{N-1} [R_{kN+i}(t)+U_{kN+i}(t)]=1. 
\end{equation}

In the stationary-state limit  Eqs. (\ref{me1}) -  (\ref{me4}) yield
\begin{equation}\label{Ri}
R_{i}=R_{0}- i p \Sigma_1, \quad \Sigma_1 = \sum_{k=1}^{\infty} R_{kN-1}+\sum_{k=1}^{\infty} U_{kN-l-1}, \quad \mbox{ for }  0 \le i \le l;
\end{equation}
\begin{equation}\label{Rl}
R_{kN+l+i}=(i+1)R_{kN+l}-i(1-p)R_{kN+l-1},  \quad  \mbox{ for }  0 \le i \le N-l; \  k \ge 0;
\end{equation}
\begin{equation}\label{Rk}
R_{kN+i}=(i+1)R_{kN}-i(1-p)R_{kN-1},  \quad  \mbox{ for }  0 \le i \le l; \  k \ge 1.
\end{equation}
Similar  results  can be obtained for the functions $U_{j}$ from Eqs. (\ref{Ri}) -  (\ref{Rk}) by changing $R\rightarrow U$, $l\rightarrow N-l$, and $\Sigma_1 \rightarrow \Sigma_2$,
\begin{equation}\label{Ui}
U_{i}=U_{0}- i p \Sigma_2, \quad \Sigma_2 = \sum_{k=1}^{\infty} U_{kN-1}+\sum_{k=0}^{\infty} R_{kN+l-1}, \quad \mbox{ for }  0 \le i \le N-l;
\end{equation}
\begin{equation}\label{Ul}
U_{kN+N-l+i}=(i+1)U_{kN+N-l}-i(1-p)U_{kN+N-l-1},  \quad  \mbox{ for }  0 \le i \le l; \  k \ge 0;
\end{equation}
\begin{equation}\label{Uk}
U_{kN+i}=(i+1)U_{kN}-i(1-p)U_{kN-1},  \quad  \mbox{ for }  0 \le i \le N-l; \  k \ge 1.
\end{equation}
The systems of Eqs. (\ref{Ri}) -  (\ref{Rk}) and Eqs. (\ref{Ui}) -  (\ref{Uk}) are coupled via the auxiliary functions $\Sigma_1$ and $\Sigma_2$.

Following the approach of Ref. \cite{morozov07}, we seek the solutions of Eqs. (\ref{Ri}) and (\ref{Rk}) in the form
\begin{equation}\label{beta}
R_{kN+i}=R_0 e^{ak} - iB(k), \quad  \mbox{ for }  0 \le i \le l,
\end{equation}
where unknown parameters $a$ and $B(k)$ are to be determined. Expressing functions $R_{kN+l}$ and $R_{kN+l-1}$ with the help of Eq. (\ref{beta}) and substituting them into Eq. (\ref{Rl}), one finds:
\begin{equation}\label{beta1}
R_{kN+l+i}=R_0 e^{ak} - lB(k)+i\left[ p R_0 e^{ak}+B(k)\{-1+p-pl\} \right] \quad  \mbox{ for } 0 \le i \le N-l.
\end{equation}
 It is important to  note that utilizing Eq. (\ref{beta}) to express $R_{kN+l-1}$ implies that $l-1\ge 0$, i.e. $l\ge 1$. Next, we use the fact that Eq. (\ref{beta1}) with $i=N-l$ should coincide with Eq. (\ref{beta}) for $i=0$, and the unknown function $B(k)$ is eliminated, leading to
\begin{equation}\label{aa}
B(k) = R_0 e^{ak}\{1-e^a+p(N-l)\}/\{l+(N-l)[1-p+pl]\}.
\end{equation}
Hence the system of Eqs. (\ref{Ri}) - (\ref{Rk}) is transformed into Eqs. (\ref{beta}) and  (\ref{beta1}) with the function $B(k)$ given by Eq. (\ref{aa}). Similarly, the solutions of the system of Eqs.  (\ref{Ui}) -  (\ref{Uk}) are sought in the following form, 
\begin{equation}\label{beta2}
U_{kN+i}=U_0 e^{ak} - i\tilde{B}(k), \quad  \mbox{ for } 0 \le i \le N-l.
\end{equation}
Then we obtain
\begin{equation}\label{beta3}
U_{kN+N-l+i}=U_0 e^{ak} - (N-l)\tilde{B}(k)+i\left[ p U_0 e^{ak}+\tilde{B}(k)\{-1+p-p(N-l)\} \right],  \mbox{ for } 0 \le i \le l
\end{equation}
with
\begin{equation}\label{aa1}
\tilde{B}(k) = U_0 e^{ak}\{1-e^a+pl\}/\{N-l+l[1-p+p(N-l)]\}.
\end{equation}

To find the unknown parameter $a$, we utilize the recurrent formula (\ref{Rk}) with the functions $R_{kN+i}$ and $R_{kN}$ expressed via Eq. (\ref{beta}) and the function  $R_{kN-1}$ found from Eq. (\ref{beta1}),
\begin{eqnarray}\label{gamma1} 
R_0 e^{ak} - iB(k)=(i+1)R_0 e^{ak}-i(1-p)\left\{R_0 e^{a(k-1)}- lB(k-1) \right.  \nonumber \\
\left.+[N-l-1]\left(pR_0 e^{a(k-1)}+B(k-1)[-1+p-pl]\right)\right\}.
\end{eqnarray}
Expressing $B(k)$ from Eq. (\ref{aa}) and introducing the parameter $x \equiv e^{a}$, one can reduce Eq. (\ref{gamma1}) into 
\begin{equation}\label{eq_x}
x^2-[2+2p(N-1)+p^2(l-1)(N-l-1)]x+(1-p)^2=0.
\end{equation}
The physically reasonable solution of Eq. (\ref{eq_x}) is $x\le 1$, that corresponds to decreasing probability of finding the particle with the distance from the last burnt bridge. Then we obtain
\begin{equation}\label{x1}
x = 1+p(N-1)+\frac{1}{2}p^2(l-1)(N-l-1)-\frac{1}{2}\sqrt{S}
\end{equation}
with
\begin{equation}\label{}
S =\left[2+2p(N-1)+p^2(l-1)(N-l-1)\right]^2-4(1-p)^2.
\end{equation}
According to Eq. (\ref{x1}), $x$ lies within the unit interval $[0,1]$ with $x=0$ for $p=1$ and $x=1$ for $p=0$. We observe that Eq. (\ref{eq_x}) and its solution Eq. (\ref{x1}) are invariant under the change $l\rightarrow N-l$, which implies that Eq. (\ref{x1}) works not only for functions $R_j$ [Eqs. (\ref{beta}), (\ref{beta1})], but also for functions $U_j$ [Eqs. (\ref{beta2}), (\ref{beta3})]. It should also be pointed out that Eq. (\ref{eq_x}) cannot be obtained from the recurrent formula (\ref{Rl}) as it leads to a trivial identity. 

In order to find the function $R_0$ from the renormalization condition (\ref{norm}), we need to eliminate the function $U_0$. This is done with the help of Eq. (\ref{Ri}) [or alternatively using Eq. (\ref{Ui})], reflecting the fact that recurrent relations for $R_j$ and those for $U_j$ are coupled through $\Sigma_1$ and $\Sigma_2$. Comparing Eq. (\ref{Ri}) with (\ref{beta}) for $k=0$, it follows that
\begin{equation}\label{sigma1}
B(0)=p\Sigma_1=p\left[\sum_{k=1}^{\infty} R_{kN-1}+\sum_{k=1}^{\infty} U_{kN-l-1}\right].
\end{equation}
While (\ref{sigma1}) can be used directly to express $U_0$ in terms of $R_0$, it is convenient to combine it with another equation, 
\begin{equation}\label{sigma}
B(0)=p\left[\sum_{k=1}^{\infty} R_{kN-1}+\sum_{k=0}^{\infty} R_{kN+l-1}\right],
\end{equation}
to obtain a simpler expression connecting $U_0$ and $R_0$. To prove the equation (\ref{sigma}), one has to express $R_{kN-1}$ and $R_{kN+l-1}$ with the help of Eqs. (\ref{beta1}) and (\ref{beta}),  correspondingly, and with the function $B(k)$ given by (\ref{aa}). This leads to proving the following expression,
\begin{equation}\label{rc}
R_0 C=\frac{pR_0}{1-x}\{1-lC+(N-l-1)[p+C(-1+p-pl)]+1-(l-1)C\},
\end{equation}
where we define a new parameter $C$ 
\begin{equation}\label{bc}
B(k)=R_0 e^{ak}C,
\end{equation}
which can be written using Eq. (\ref{aa}),
\begin{equation}\label{cc}
C=\frac{1-x+p(N-l)}{l+(N-l)[1-p+pl]}.
\end{equation}
It can be shown that Eq. (\ref{rc}) is equivalent to equation (\ref{eq_x}) for $x$, and therefore (\ref{rc}) is correct, and so is (\ref{sigma}). Then comparing Eqs. (\ref{sigma1}) and (\ref{sigma}) it follows that
\begin{equation}\label{ur1}
\sum_{k=1}^{\infty} U_{kN-l-1}=\sum_{k=0}^{\infty} R_{kN+l-1}. 
\end{equation}
Expressing $U_{kN-l-1}$ and $R_{kN+l-1}$ according to Eqs. (\ref{beta2}) and (\ref{beta}), we obtain
\begin{equation}\label{u11}
U_{kN-l-1}=U_0 x^{k-1}[1-(N-l-1)\tilde{C}], 
\end{equation}
where we defined another parameter $\tilde{C}$,
\begin{equation}\label{bc1}
\tilde{B}(k)=U_0 e^{ak}\tilde{C},
\end{equation}
which is equal to [see Eq. (\ref{aa1})]
\begin{equation}\label{cc1}
\tilde{C}=\frac{1-x+pl}{N-l+l[1-p+p(N-l)]}.
\end{equation}
Combining these results with 
\begin{equation}\label{r11}
R_{kN+l-1}=R_0x^k[1-(l-1)C]
\end{equation}
and summing over the index $k$ in (\ref{ur1}) yields 
\begin{equation}\label{u0r0}
U_0=\frac{1-(l-1)C}{1-(N-l-1)\tilde{C}}R_0.
\end{equation}
Eq. (\ref{u0r0}) is the relation between $U_0$ and $R_0$ that is needed in order to determine the function $R_0$. It should be noted that the same result can be obtained by using Eq. (\ref{Ui}), i.e. $\tilde{B}(0)=p\Sigma_2$. 

Now we can find $R_0$ from the normalization condition (\ref{norm}), which can be rewritten as
\begin{equation}\label{norm1}
\sum_{k=0}^{\infty} \left[\sum_{i=0}^{l-1} R_{kN+i}+\sum_{i=0}^{N-l-1} R_{kN+l+i}+\sum_{i=0}^{N-l-1} U_{kN+i}+\sum_{i=0}^{l-1} U_{kN+N-l+i}\right]=1. 
\end{equation}
We also rewrite functions $R_{kN+i}$, $R_{kN+l+i}$ and $U_{kN+i}$, $U_{kN+N-l+i}$ according to Eqs. (\ref{beta}), (\ref{beta1}),  (\ref{beta2}) and (\ref{beta3}), with functions  $B(k)$ and  $\tilde{B}(k)$ given by expressions (\ref{aa}) and (\ref{aa1}),
\begin{eqnarray}
R_{kN+i}&=&R_0 x^k[1-iC]; \label{r12} \\
R_{kN+l+i}&=&R_0 x^k[1-lC+i\{p+C(-1+p-pl)\}]; \label{r15} \\
U_{kN+i}&=&U_0 x^k[1-i\tilde{C}]; \label{u12} \\
U_{kN+N-l+i}&=&U_0 x^k[1-(N-l)\tilde{C}+i\{p+\tilde{C}(-1+p-p[N-l])\}],\label{u15}
\end{eqnarray}
where $C$ and $\tilde{C}$ are provided by (\ref{cc}) and (\ref{cc1}). We note that Eqs. (\ref{u12}) and (\ref{u15}) correspond to Eqs. (\ref{r12}) and (\ref{r15}) with $l\rightarrow N-l$ change (since the same correspondence exists between $C$ and $\tilde{C}$). The normalization (\ref{norm1}) implies that
\begin{eqnarray}\label{norm2}
\frac{R_0}{1-x}\left\{l-\frac{1}{2}l(l-1)C+(N-l)(1-lC) \right. \nonumber \\
\left.+\frac{1}{2}(N-l)(N-l-1)[p+C(-1+p-pl)]\right\} \nonumber \\
+\frac{U_0}{1-x}\left\{N-l-\frac{1}{2}(N-l)(N-l-1)\tilde{C}+l(1-(N-l)\tilde{C})
\right. \nonumber \\
\left. +\frac{1}{2}l(l-1)[p+\tilde{C}(-1+p-p[N-l])]\right\} =1
\end{eqnarray}
and, introducing parameter $B$ such that $U_0=BR_0$, we obtain from Eq. (\ref{u0r0})
\begin{equation}\label{BB}
B=\frac{1-(l-1)C}{1-(N-l-1)\tilde{C}}.
\end{equation}
Then  from (\ref{norm2}) the expression for $R_0$ can be derived,
\begin{eqnarray}\label{r00}
R_0=2(1-x)/ \left\{  2l-l(l-1)C+2(N-l)(1-lC) \right.  \nonumber \\
+(N-l)(N-l-1)[p+C(-1+p-pl)] \nonumber \\
+B \left\{ 2(N-l)-(N-l)(N-l-1)\tilde{C}+2l(1-(N-l)\tilde{C}) \right.  \nonumber \\
\left. \left. +l(l-1)[p+\tilde{C}(-1+p-p(N-l))] \right\} \right\}.
\end{eqnarray}

To find the mean velocity of the dimer molecular motor  we utilize the formula discussed in \cite{morozov07},
\begin{equation}
V = \sum\limits_{j=0}^{\infty}(u_j-w_j)R_j+\sum\limits_{m=0}^{\infty}(\tilde{u}_m-\tilde{w}_m)U_m.
\end{equation}
The physical meaning of this expression is the fact that the velocity can be viewed as a sum of the terms for each possible state of the system as shown in Fig. 8. Due to the cancellation of forward and backward rates on all sites except the origin, we obtain the  expression,
\begin{equation}\label{VV}
V =R_0+U_0=(1+B)R_0,
\end{equation}
with functions $R_0$ and $B$ given by (\ref{r00}) and (\ref{BB}). Eq. (\ref{VV}) provides the exact formula for the dimer's velocity $V(l,N,p)$ (or $V(l,c,p)$) in the continuous-time forward BBM.

It can be shown that the derived expression for the velocity agrees with previously obtained exact formula in the special cases. First, for $p=1$ the parameters needed to calculate $V$ are the following,
\begin{equation}\label{}
x=0, \quad C=\frac{1}{l}, \quad \tilde{C}=\frac{1}{N-l}, \quad B=\frac{N-l}{l},
\end{equation}
and Eq. (\ref{VV}) yields
\begin{equation}\label{}
V(l,N,p=1)=\frac{2N}{l(l+1)+(N-l)(N-l+1)},
\end{equation}
which reproduces the result (\ref{vel1}) obtained for $p=1$ by using  Derrida's method. Another special case is when the bridges are shifted by $l=N/2$ lattice spacings with $0<p\le 1$ (discussed in the preceding section). We checked  numerically for complete ranges of $c$ and $p$ (with various fixed $p$ and $c$ values correspondingly) and in this way verified that Eq. (\ref{VV}) produces the correct results.

The formula (\ref{VV}) is derived for non-zero shifts, $l=1,2,\cdots,N-1$, as discussed above. We now consider the case of $l=0$, when a dimer crosses pairs of bridges on the upper and lower tracks at once. Instead of trying to apply the kinetic scheme in Fig. 8 to this case, we observe that a pair of bridges on parallel lattices crossed simultaneously by a dimer is equivalent to a single ``effective'' bridge on the single track but with a different burning probability $p_{eff}$. Thus the motion of the dimer along two parallel tracks can be viewed as hopping of a new effective particle in single-lattice BBM with effective burning probability.  The velocity for a single random walker in the continuous-time forward BBM  was found in \cite{morozov07},
\begin{equation}\label{vpeff}
V(N,p_{\rm{eff}}) = \frac{2p_{\rm{eff}}}{\sqrt{p_{\rm{eff}}^2(N-1)^2 + 4p_{\rm{eff}}N}}.
\end{equation}
The effective burning probability can be found from the following arguments. Both bridges can be burnt simultaneously with the probability of $p^{2}$, or only one bridge can be destroyed at each lattice, producing 
\begin{equation}\label{}
p_{\rm{eff}}=2p(1-p)+p^2=2p-p^2,
\end{equation}
and Eq. (\ref{vpeff}) leads to the following expression,
\begin{equation}\label{vl0}
V(l=0,N,p) = \frac{2(2p-p^2)}{\sqrt{(2p-p^2)^2(N-1)^2 + 4(2p-p^2)N}}.
\end{equation}
We note that, as expected, $p_{\rm{eff}}>p$ for $0<p<1$, and  $p_{\rm{eff}}=p$ if $p=0$ or 1. Thus for $0<p<1$, the dimer's velocity $V(l=0,N,p)$ always exceeds that of the monomer at the same conditions. It is also interesting to note that Eq. (\ref{vl0}) can be obtained by direct substituting $l=0$ into expression (\ref{VV}) that was obtained for $l \ge 1$ cases.

The result (\ref{VV}) allows us to compare  velocities of the dimer and the monomer molecular motors in BBM for the entire range of burning probabilities. The ratio of velocities at different conditions is plotted in Figs 9-11. The dimer's velocity $V_2(l,c,p)$ is calculated from Eq. (\ref{VV}), while the velocity of the monomer $V_1(c,p)$ is given by Eq. (\ref{v1}). As shown in Fig. 9a, the ratio $V_{2}/V_{1}$ shows a non-monotonic behavior as a function of the bridge concentration with a maximum that approaches 2 (from below in the limit of very large shift distances as long as the burning probability $p$ is not too small. Also the ratio of velocities  as a function of the shift distance $l$ (for fixed $c$) has a maximum at $lc=1/2$: see Fig. 9b. This behavior is similar to the case of $p=1$: compare Fig. 3 and Fig. 9. The effect of the burning probability on velocities is illustrated in Fig. 10. Increasing the burning probability leads to the increase in the ratio of the dimer's and the monomer's velocities for the intermediate values of $c$ and $l$, while for very large and very small concentrations, as well as very small and very large shifts, the trend is reversed. A special case of $l=0$ is illustrated in Fig. 11. In this case the maximum value of $V_2/V_1$ is reached at $c=1$ (when every site is a potential bridge), and for all concentrations $V_2/V_1$ increases as $p$ decreases. The ratio $V_2/V_1$ varies from 1 for $p=1$ to $\sqrt{2}$ in the $p\rightarrow 0$ limit (provided that $p\ll c$).

\begin{figure}[tbp]
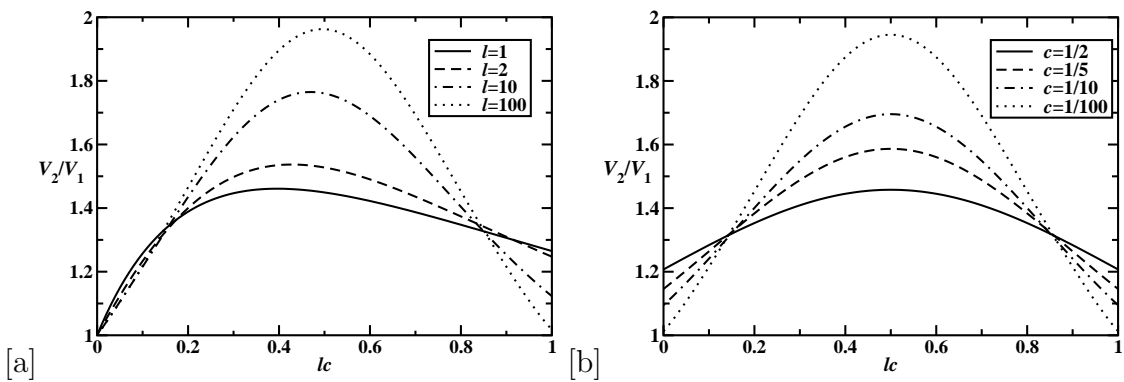

\centering
[a]\includegraphics[scale=0.28,clip=true]{Fig9a.eps}
[b]\includegraphics[scale=0.28,clip=true]{Fig9b.eps}
\caption{(a) Ratio of dimer's and monomer's velocities as a function of the concentration of bridges $c$ for the fixed value of the shift $l$ for $p=0.4$.  (b) Ratio of dimer's and monomer's velocities as a function of the shift $l$ for the fixed concentration of bridges $c$ for $p=0.5$.}
\end{figure}

\begin{figure}[tbp]
\centering
[a]\includegraphics[scale=0.28,clip=true]{Fig10a.eps}
[b]\includegraphics[scale=0.28,clip=true]{Fig10b.eps}
\caption{(a) Ratio of dimer's and monomer's velocities as a function of the concentration of bridges $c$ for the fixed value of the shift $l=5$ at different burning probabilities.  (b) Ratio of dimer's and monomer's velocities as a function of the shift $l$ for the fixed concentration of bridges $c=0.1$ at different burning probabilities.}
\end{figure}

\begin{figure}[tbp]
\centering
\includegraphics[scale=0.28,clip=true]{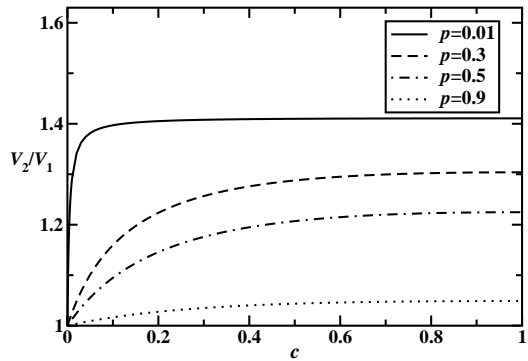}
\caption{Ratio of dimer's and monomer's velocities as a function of the concentration of bridges $c$ for the fixed value of the shift $l=0$ at different burning probabilities.}
\end{figure}

\section{Discussions}

Our theoretical calculations show that the rigid dimer molecular motor always moves faster than the monomer in continuous-time BBM. However, the ratio of velocities never exceeds 2 (see Figs. 3, 6a, 9, 10 and 11). The behavior of the diffusion constant is more complex, as indicated in Figs. 4 and 6b, although the ratio $D_{2}/D_{1}$ is always between 0.9 and 1.31.  It is interesting to note that for some sets of parameters fluctuations of the dimer are smaller than fluctuations of the monomer particle at the same conditions. This observation can be understood using the following arguments. The fluctuations of the particle in BBM are reduced near the already burned bridges because the particle cannot cross them. In the case of the dimer the effective concentration of weak links is larger than for the monomer, and the dimer molecular motor more frequently encounters hard wall from the left, leading to decrease in the dispersion for some range of parameters.

More interesting is the behavior of effective forces exerted by the dimer particles in BBM presented in Figs. 5 and 7. The dimer always generates larger force than the monomer particle, but in most cases the ratio $F_{2}/F_{1}$ is less than 2. The situation is different for the special case of $l=1$ and $c=1/2$ with the burning probability $p$ close or equal to 1. Here the effective force exerted by the dimer becomes very large. This behavior can be easily understood for $p=1$ burnt-bridge models. In this case the motion of the dimer on two parallel lattices can be viewed as the motion of the effective  particle on a single lattice with larger concentration of bridges - as shown in Fig. 2. Because $N=1/c=2$ and $l=1$ every link in the effective single lattice is a potential bridge. Then the random walker always move forward, and the backward transition rate is zero, leading to infinite effective exerted forces for the dimer molecular motor [see Eq.(\ref{Fs})].

Computer simulations of the dynamics of clusters of collagenase proteins \cite{saffarian06} predict that a rigid dimer can generate large forces. In collagen fibers the cleavage sites are separated by the distance $\Delta=300$ nm, and fibers are shifted by approximately 60 nm from each other \cite{lodish_book,saffarian04}. Although the step-size for the collagen motor protein is unknown, one can reasonably take it as $\simeq 10$ nm, the value typical for most known motor proteins and consistent with the sizes of these enzymes \cite{lodish_book,AR}. Then it would correspond to $N \simeq 30$ (or $c \simeq 1/30$), and $l\simeq 6$. Experimental and theoretical estimates of the burning probability are between 10 and 30 $\%$ \cite{saffarian04,morozov07}. However, even if one assumes that $p=1$ at these conditions our analytical calculations [from Eq. (\ref{f2})] produce  $F_{2}\simeq 0.045$ pN, which is still significantly smaller than predicted from Monte Carlo computer simulations for dimers  \cite{saffarian06}. Thus our analysis suggests that the dimer collagenases, in contrast with earlier claims \cite{saffarian06}, probably cannot produce stall forces comparable to ATP-driven motor proteins.

\section{Summary and conclusions}

The dynamics of rigid dimer particles moving along parallel molecular tracks in BBM is investigated theoretically using several analytical methods. When the probability of burning is equal to 1, we utilized Derrida's method \cite{derrida83} for calculating explicitly dynamic properties of the system, and it allowed us to compute the effective forces exerted by the dimer molecular motor. For $p<1$ the situation when the bridges on the parallel tracks are shifted by half of the period, i.e., $l=N/2$, is also investigated in detail by mapping the system with two tracks into the motion along the single lattice with the distance between bridges being $N/2$.  More generally, for $ p \le 1$ and arbitrary shifts between the tracks, our analysis produced exact analytical expressions for the velocity of the dimer. 

 The dynamic properties and the force generated by the dimer are compared with corresponding properties for the monomer molecular motor. In all cases  it was found that the ratio of the dimer's and the monomer's velocities  can never exceed 2 (which is the limit in case of small concentration of bridges $c$ or large values of the shift $l$). We found that the force generated by the dimer is substantially larger than that generated by the monomer only for a special case of  $c=1/2$ and  $l=1$. For other sets of parameters the force exerted by the dimer is at most only slightly more than twice as large as that exerted by the monomer. This occurs for both $p=1$ and $p<1$ (with $l=N/2$), in the latter case the significant difference between the forces generated by the dimer and the monomer at $c=1/2$ is present only if the burning probability $p$ is close to 1. This result is in contradiction with earlier theoretical claims based on Monte Carlo computer simulations \cite{saffarian06}. These theoretical predictions are explained by the increase in the effective concentration of bridges.

Although our theoretical approach provides exact and explicit expressions for dynamic properties of dimer molecular motors in BBM, there are several features of the system that should be included in order to obtain more realistic description of dynamics of motor proteins that interact with their tracks. It was assumed that two motor proteins are tightly bound in the dimer, while in the cell the enzymes must have some flexibility. In addition, the interaction between motor proteins in the cluster was neglected, although it might lead to the increased dynamic efficiency, as was discussed earlier for different motor protein systems \cite{stukalin05}. Also, it is not clear what effect on the dynamics of molecular motors will have random or non-uniform distributions of bridges. It is important to investigate further these features theoretically and experimentally to uncover the fundamental mechanisms of motor protein transport. 

\ack
 
The support  from the Welch Foundation (under Grant No. C-1559), and from the US National Science Foundation (grants CHE-0237105 and NIRT ECCS-0708765) is gratefully acknowledged. The authors thank M.N. Artyomov for valuable discussions and suggestions.

\end{document}